\documentclass[epj]{svjour}
\usepackage{graphics}
\usepackage{amssymb}
\def\q1{{q^{-1}}}
\def\qq1{{q-q^{-1}}}

\begin{document}
\title{Classical and quantum $q$-deformed physical systems}
\author{A. Lavagno\inst{1,2}, A.M. Scarfone\inst{1,3} \and P. Narayana Swamy\inst{4}
}                     
\institute{Dipartimento di Fisica, Politecnico di Torino,
Italy \and INFN-Istituto Nazionale di Fisica Nucleare,
Sezione di Torino, Italy. \and INFM/CNR Istituto Nazionale
di Fisica della Materia, Sezione di Torino, Italy. \and
Southern Illinois University, Edwardsville, IL 62026, USA.}
\mail{antonio.scarfone@polito.it}
\date{Received: date / Revised version: date}
%
\abstract{ On the basis of the non-commutative $q$-calculus, we
investigate a $q$-deformation of the classical Poisson bracket in
order to formulate a generalized $q$-deformed dynamics in the
classical regime. The obtained $q$-deformed Poisson bracket appears
invariant under the action of the $q$-symplectic group of
transformations. In this framework we introduce the $q$-deformed
Hamilton's equations and we derive the evolution equation for some
simple $q$-deformed mechanical systems governed by a scalar
potential dependent only on the coordinate variable. It appears that
the $q$-deformed Hamiltonian, which is the generator of the equation
of motion, is generally not conserved in time but, in
correspondence, a new constant of motion is generated. Finally, by
following the standard canonical quantization rule, we compare the
well known $q$-deformed Heisenberg algebra with the algebra
generated by the $q$-deformed Poisson bracket.
} 
\authorrunning{A. Lavagno et al.}
\maketitle



\section{Introduction}

Quantum algebra and quantum groups arise as the underlying
mathematical structure in several physical phenomena. It has been
shown that such a formalism can play an important role in
conformal field theory, exact soluble models in statistical
physics \cite{Baxter}
and in a wide range of applications, from cosmic strings and black
holes to  solid state physics problems \cite{wil,lerda,alva}.

Many physical applications have been investigated on the basis of
 the $q$-deformation of the Heisenberg algebra
\cite{Wess,Hinterding,Cerchiai,meljanac,iida}. For
instance,  $q$-deformed Schr\"odinger equations have been
proposed in the literature \cite{Zhang,Micu} and
applications to the study of $q$-deformed version of the
hydrogen atom and of the quantum harmonic oscillator
\cite{Finkelstein,Lorek,Kwek} have been presented. In
Ref. \cite{Celeghini1} the Weyl-Heisenberg algebra has been
studied in the framework of the Fock-Bargmann
representation allowing a rigorous treatment of the
squeezed states,
lattice quantum mechanics and Bloch functions.

The theory of the $q$-deformed harmonic oscillator, ba\-sed on the
construction of SU$_q$(2) algebra of $q$-deformed commutation or
anticommutation relations between creation and annihilation
operators \cite{bie,mac,gruver}, has opened the possibility of
studying intermediate $q$-boson and $q$-fermion statistical
behavior \cite{chia,song}. A kinetic approach of this problem,
within a semiclassical treatment, was presented in Ref. \cite{Lavagno}.
Moreover, it has  recently lead to the application of $q$-calculus
in the construction of generalized statistical mechanics where
nonextensivity properties can arise from the $q$-deformed theory
\cite{gellmann}. Along these lines, a generalized
thermostatistics based on the formalism of  $q$-calculus has been
formulated in Ref. \cite{pre}, whereas in Ref. \cite{Abe2} a
$q$-deformed entropy was applied to
study a gas of $q$-deformed bosons.

In the recent past, some tentative approaches to construct a
$q$-deformed version of  classical mechanics have been investigated.
The primary motivation for such a stu\-dy is to understand the
origin of the $q$-deformed Heisenberg algebra which forms  the
basis of the deformed quantum mechanics. In other words, we
can ask: how does one introduce a $q$-deformed algebraic structure
for the quantum plane coordinates, such that, after canonical
quantization, the $q$-deformed Heisenberg algebra follows? This
question was investigated, for instance, in Ref. \cite{Chaichian} where
the quasi-classical limit of the $q$-oscillator has been discussed
and a $q$-deformed version of the Poisson bracket (PB) was derived
in terms of variables of the quantum plane. A related question is
how one can  describe the dynamical evolution of a classical
object existing in such a $q$-deformed quantum plane.
In Ref. \cite{Klimek} the author has  presented a tentative formulation to
construct  $q$-deformed classical mechanics based on the
introduction of a $q$-La\-gran\-gian and a $q$-Hamiltonian where the
 equations of motion are derived from the $q$-deformed analog of the
Euler-Lagrange equation. A similar formulation was presented in Ref. \cite{Lukin}
by introducing a deformed phase-space based on
the elliptic algebra with different deformed parameters for the
space coordinates and the momenta.

Finally, among the many motivations for the  study of a
$q$-deformed generalization of  classical mechanics, it is
important to mention the relevance of  symmetries in physics
\cite{Chung}. In this respect, quantum groups give a new symmetric
aspect for the resolution of classical problems. Along these
lines,  Ref. \cite{Avan1} deals with the  investigation  of a set
of Poisson algebra structures obtained  from the elliptic quantum
algebra. It has been shown that the resulting Poisson structure
contains the $q$-deformed Virasaro algebra which plays a central
role in the resolution of several integrable systems both in
quantum mechanics and
statistical mechanics.

Recently, in Ref. \cite{Noi}, a possible definition of
$q$-defor\-med PB has been derived by requiring that it  be
invariant under the action of the $q$-deformed symplectic group
Sp$_q$(1), in analogy with the classical (undeformed) case, where
PB are invariant under the action of the symplectic
group Sp(1).
In this paper we start by considering a $q$-deformed version of PB
previously introduced but following another approach, and in
some sense, a more systematic derivation instead of
 the one adopted in Ref. \cite{Noi}.

In the commutative classical mechanics, the PB between two
functions $f(x,\,p)$ and $g(x,\,p)$ can be defined through
the contraction of the corresponding Hamiltonian fields
$X_f$ and $X_g$ with the canonical symplectic form
$\omega=dx\wedge dp$. In the same manner, we can define a
$q$-deformed version of PB between two $q$-functions $\hat
f(\hat x,\,\hat p)$ and $\hat g(\hat x,\,\hat p)$, defined
on the noncommutative $q$-plane, through the contraction of
the corresponding $q$-Hamiltonian fields $\hat X_{\hat f}$
and $\hat X_{\hat g}$ with the $q$-deformed canonical
symplectic form $\hat\omega=d\hat x\wedge d\hat p$
(throughout this paper we denote by a hat the elements
belonging to the $q$-algebra to distinguish them from the
element belonging to the ordinary commutative algebra). We
then attempt  to formulate, by means of the $q$-deformed
Hamilton's equations, a $q$-deformed classical mechanics
describing the time evolution of a mechanical system.
Finally, in analogy with  standard canonical quantization
method, where the (undeformed) PB between canonically
conjugate variables $x$ and $p$ are replaced by the
(undeformed) commutator of the corresponding quantum
operators $\overline x$ and $\overline p$, we discuss the
scenario of a possible canonical quantization in the
$q$-deformed framework, putting in correspondence the
algebra generated by the $q$-deformed PB between $\hat x$
and $\hat p$ with the well known algebra generated by the
$q$-deformed commutator of the corresponding $q$-deformed
operators $\tilde x$ and $\tilde p$.

Our paper is organized as follows. After a brief review of the
derivation of the standard PB in the formalism of the exterior
calculus, presented in Sec. 2, we introduce, in Sec. 3, the
$q$-commutative phase-space and recall the definition of its
$q$-calculus. On the basis of the previous sections, we will be able
to obtain the most original result of our paper: the introduction of
a $q$-deformed PB in Sec. 4 and the formulation of a possible
$q$-deformed mechanics by means of $q$-Hamilton's equations is
presented in Sec. 5. In Sec. 6, starting from the $q$-deformed PB we
explain a possible derivation of the $q$-deformed Heisenberg
algebra. Finally the conclusions are presented in Sec. 7.

\section{Poisson bracket in the commutative phase space}
We begin by recalling briefly the derivation of the standard PB in
the formalism of the exterior calculus, referring to the relevant
literature for the
details \cite{Abrahams,Jose}.

Let us consider the real plane $I\!\!R^2$ generated by the
commutative coordinates $x^1\equiv x$ and $ x^2\equiv p$
and introduce the associative algebra ${\mathcal A}=$
Fun$(I\!\!R^2)$ of the functions on $I\!\!R^2$ freely
generated by the elements $x$ and $p$.\\ The tangent space
$T{\mathcal A}$ on $\mathcal A$ is generated by the vectors
$\partial_1\equiv\partial_x=\partial/\partial x$ and
$\partial_2\equiv\partial_p=\partial/\partial p$ which are
linear operators i.e.,
$\partial_i(\lambda\,f+\mu\,g)=\lambda\,\partial_if
+\mu\,\partial_ig$ and satisfying the Leibniz rule
$\partial_i(x^j\,f)=\delta^{\scriptscriptstyle
j}_i\,f+x^j\,\partial_if$ with $i,\,j=1,\,2$, where
$f(x,\,p)$ and $g(x,\,p)$ are smooth functions on $\mathcal
A$ and $\lambda,\,\mu$ are ordinary commutative numbers
($C$-numbers). Any vector $v$ can be spanned on the base of
$\partial_i$ as $v=f^1\,\partial_x+f^2\,\partial_p$, where
$f^i\in{\mathcal A}$.

In the same way we introduce the cotangent space
(one-forms) $T^\ast{\mathcal A}$, generated by the elements
$dx$ and $dp$. Any one-form $\omega$ can be spanned on this
base as $\omega=dx\,g_1+dp\,g_2$, where $g_i\in{\mathcal
A}$.

Higher order forms are constructed by means of the differential
operator $d$ which takes $k$-forms into $(k+1)$-forms. In
particular starting from a function $f\in {\mathcal A}$ (0-form),
its differential is a 1-form
\begin{equation}
df\equiv dx\,\partial_xf+dp\,\partial_pf \ ,\label{d}
\end{equation}
and, starting from a 1-form $\omega=dx\,g_1+dp\,g_2$ we obtain the
2-form
\begin{equation}
d\omega\equiv dx\wedge
dp\,\left(\partial_pg_2-\partial_xg_1\right) \ ,\label{2f}
\end{equation}
where we have introduced the exterior product between two
1-forms, that is linear
$\left(\lambda\,\omega_{_1}+\mu\,\omega_{_2}\right)\wedge\omega_{_3}=\lambda\,\left(
\omega_{_1}\wedge\omega_{_3}\right)\\+\mu\,\left(
\omega_{_2}\wedge\omega_{_3}\right)$, associative
$\left(\omega_{_1}\wedge\omega_{_2}\right)\wedge\omega_{_3}=
\omega_{_1}\wedge\left(\omega_{_2}\wedge\omega_{_3}\right)$
and skew-symmetric
$\omega_{_1}\wedge\omega_{_2}=-\omega_{_2}\wedge\omega_{_1}$.\\
The $d$ operator fulfills the following main properties:
$d(\lambda\,\omega\\+\mu\,\omega^\prime)=\lambda\,d\omega+\mu\,d\omega^\prime$
(linearity); $d(f\,g)=df\,g+f\,dg$ (Leibniz rule);
$d\lambda=0$; $d(d\omega)=0$; and
$d(\omega^{(k)}\wedge\omega^{(p)})=d\omega^{(k)}\wedge\omega^{(p)}
+(-1)^k\,\omega^{(k)}\wedge d\omega^{(p)}$ where $k$ and
$p$ are the degree of $\omega^{(k)}$ and $\omega^{(p)}$,
respectively.

Finally, we introduce the contraction operator $i_v(\omega)$, in
the axiomatic way, through its main propriety listed below,
\begin{eqnarray}
\nonumber
&&i_v(f)=0 \ , \\
\nonumber
&&i_{\partial_i}(dx^j)=\delta_i^{\,\,j} \ ,\\
&&i_{(\lambda\,f\,\partial_i+\mu\,g\,\partial_j)}(\omega)=\lambda\,f\,i_{\partial_i}(\omega)
+\mu\,g\,i_{\partial_j}(\omega) \ ,\label{i}\\
\nonumber
&&i_v\,(\lambda\,dx^i\,f+\mu\,dx^j\,g)=\lambda\,i_v(dx^i)\,f+\mu\,i_v(dx^j)\,g \ ,\\
\nonumber &&i_{\partial_i}\,(dx^1\wedge
dx^2)=\delta_i^1\,dx^2-\delta^2_i\,dx^1 \ .
\end{eqnarray}
In order to derive the Poisson bracket and their algebra
within this formalism we begin by introducing the
symplectic form
\begin{equation}
\omega=dx\wedge dp \ ,\label{csym}
\end{equation}
and define the Hamiltonian vector field $X_f$, associated with a
function $f\in{\mathcal A}$, through the relation
\begin{equation}
i_{_{X_f}}\,(\omega)=d\,f \ .\label{ham}
\end{equation}
>From Eqs. (\ref{csym}) and (\ref{ham}), taking into account
 the properties of the contraction operator (\ref{i}),
it follows that, for any function $f$, the corresponding
Hamiltonian vector field $X_f$ assumes the expression
\begin{equation}
X_f\equiv
\partial_pf\,\partial_x-\partial_xf\partial_p \ .\label{xh}
\end{equation}
As a consequence, we can write the Poisson bracket between the two functions
$f,\,g\in {\mathcal A}$ through the relation
\begin{equation}
\Big\{f,\,g\Big\}\equiv
i_{X_g}(df)=i_{X_g}\,i_{X_f}(\omega) \ .\label{pb}
\end{equation}
It is easy  to show, accounting for Eq. (\ref{xh}), that Eq.
(\ref{pb}) can be written in the usual form
\begin{equation}
\Big\{f,\,g\Big\}=\partial_xf\,\partial_pg-\partial_pf\partial_xg
\ .\label{pb2}
\end{equation}
We recall the main properties of the PB defined through Eq.
(\ref{pb}) which are: bi-linearity
$\{\lambda\,f+\mu\,g,\,h\}=\lambda\,\{f,\,h\}+\mu\,\{g,\,h\}$ and
$\{f,\,\lambda\,g+\mu\,h\}=\lambda\,\{f,\,g\}+\mu\,\{f,\,h\}$,
skew symmetry $\{f,\,g\}=-\{g,\,f\}$ and the Jacobi identity $ \\
\{f,\,\{g,\,h\}\}+\{g,\,\{h,\,f\}\}+\{h,\,\{f,\,g\}\}=0$.

We remark that Eq. (\ref{pb2}) can  also be expressed in
the form
\begin{equation}
\Big\{f,\,g\Big\}=\partial_if\,J^{ij}\,\partial_jg \
,\label{poisson1}
\end{equation}
(repeated indexes are summed over), where we have
introduced the structure functions
\begin{equation}
J^{ij}=\Big\{x^i,\,x^j\Big\} \ ,\label{j}
\end{equation}
which can be arranged in $2\times2$ matrix $J$. Taking into
account the expression of the symplectic form (\ref{csym}), we
easily recognize that $J$ is the symplectic unity, with entries
$J^{ij}=\epsilon^{ij}$, where $\epsilon^{12}=-\epsilon^{21}=1$.\\
It is easy to show that Eq. (\ref{poisson1}) is invariant
in form under a symplectic transformation Sp(1) on the
plane $I\!\!R^2$
\begin{equation}
\Big\{f,\,g\Big\}\to\Big\{f^\prime,\,g^\prime\Big\}=
\partial_i^\prime f^\prime\,J^{ij}\,\partial_j^\prime g^\prime
\ ,
\end{equation}
with $f^\prime=f(x^\prime,\,p^\prime)$ and
$g^\prime=g(x^\prime,\,p^\prime)$ and
\begin{equation}
x^i\rightarrow {x^\prime}^i= x^j\,T_j^{\,\,i} \ ,
\end{equation}
where $T_j^{\,\,i}$, the entries of a matrix $T\in\;$Sp(1),
satisfies the symplectic relation
\begin{equation}
T_m^{\,\,i}\,J^{mn}\,T_n^{\,\,j}=J^{ij} \ .\label{cttc}
\end{equation}

\section{$q$-commutative differential calculus}

In order to generalized the PB in the framework of a $q$-deformed theory,
we review the basic properties of the $q$-commutative differential calculus.

The real quantum plane $\widehat{I\!\!R}^2$ is generated by
the $q$-commutative element $\hat x^1\equiv\hat x$ and
$\hat x^2\equiv\hat p$, obeying the relation
\cite{Reshetikhin}
\begin{equation}
\hat p\,\hat x=q\,\hat x\,\hat p \ , \label{qplane}
\end{equation}
which is invariant under the action of  Gl$_q$(2) transformations
and $q$ is the real deformation parameter. We denote by
$\widehat{\mathcal A}=$ Fun$(\widehat{I\!\!R}^2)$ the associative
algebras freely generated by the elements $\hat
x$ and $\hat p$.\\
In analogy with the commutative case, we define the
$q$-tangent space $T\widehat{\mathcal A}$ \cite{Aschieri},
generated by the $q$-derivatives
$\hat\partial_1\equiv\hat\partial_x$ and
$\hat\partial_2\equiv\hat\partial_p$, whose action on the
generators $\hat x^i$ is defined as \cite{Wess}
\begin{equation}
\hat\partial_i\,\hat x^j=\delta_i^{\,\,j} \ .\label{der}
\end{equation}
They are linear operators satisfying $\hat\partial_i(\lambda\,\hat
f+\mu\,\hat g)=\lambda\,\hat\partial_i\hat f
+\mu\,\hat\partial_i\hat g$ which fulfill the $q$-Leibniz rule
\begin{eqnarray}
\nonumber &&\hat\partial_p\,\hat p=1+q^2\,\hat
p\,\hat\partial_p+(q^2-1)\,\hat x\,\hat\partial_x \ ,\\
&&\hat\partial_p\,\hat x=q\,\hat x\,\hat\partial_p\ ,\label{qco}\\
\nonumber
&&\hat\partial_x\,\hat p=q\,\hat p\,\hat\partial_x\ ,\\
\nonumber &&\hat\partial_x\,\hat x=1+q^2\,\hat
x\,\hat\partial_x \ ,
\end{eqnarray}
leading to the $q$-commutative derivative
\begin{equation}
\hat\partial_p\,\hat\partial_x=q^{-1}\,\hat\partial_x\,\hat\partial_p\ .
\end{equation}

Let us now introduce the $q$-cotangent space
$T^\ast\widehat{\mathcal A}$, generated by the elements $d\hat x$
and $d\hat p$. Any $q$-deformed 1-form $\hat\omega$ can be spanned
on this basis as $\hat\omega=d\hat x\,\hat
g_1+d\hat p\,\hat g_2$ where $\hat g_i\in\widehat{\mathcal A}$.
We observe that, $d\hat x^i\,\hat f\not=\hat f\,d\hat x^i$, due to
the $q$-commutative structure of the $q$-calculus. Nevertheless,
accounting for the following relations
\begin{eqnarray}
\nonumber &&\hat p\,d\hat p=q^2\,d\hat p\,\hat p \
,\\
&&\hat x\,d\hat p=q\,d\hat p\,\hat x \
,\label{qd}\\
\nonumber &&\hat p\,d\hat x=q\,d\hat x\,\hat
p+(q^2-1)\,d\hat p\,\hat x\
,\\
\nonumber &&\hat x\,d\hat x=q^2\,d\hat x\,\hat x \ ,
\end{eqnarray}
any $q$-deformed 1-form is well defined and admits a
unique, left or right, expansion: $\hat\omega=\hat f^{\rm
L}_i\,d\hat x^i=d\hat x^i\,\hat f_i^{\rm R}$ where the
quantities $\hat f^{\rm R}_i$ and $\hat f^{\rm L}_i$ can be
obtained each from the other by means of Eqs. (\ref{qd}).

Through the action of the operator $d$ we can construct
higher order $q$-deformed forms. In particular, the
differential of a $q$-function $\hat f\in \widehat{\mathcal
A}$ is given by
\begin{equation}
d\hat f\equiv d\hat x\,(\hat\partial_x\hat f)^{\rm R}+d\hat
p\,(\hat\partial_p\hat f)^{\rm R} \ ,\label{qd1}
\end{equation}
which, by means of Eqs. (\ref{qd}), can be written equivalently
in the (left) form
\begin{equation}
d\hat f\equiv (\hat\partial_x\hat f)^{\rm L}\,d\hat
x+(\hat\partial_p\hat f)^{\rm L}\,d\hat p \ .\label{qd2}
\end{equation}
We recall that in the framework of the $q$-calculus, the
operator $d$ fulfills the same formal properties as in
standard calculus. The elements $d\hat x$ and $d\hat p$
satisfy the relations
\begin{eqnarray}
\nonumber
&&d\hat p\wedge d\hat p=0 \ ,\\
&&d\hat p\wedge d\hat x=-q^{-1}\,d\hat x\wedge d\hat p \
,\label{d1d2}\\
\nonumber &&d\hat x\wedge d\hat x=0 \ ,
\end{eqnarray}
where, the exterior product is still linear and associative.
Finally, we introduce the contraction operator $i_v(\omega)$
between $q$-deformed vectors and $q$-deformed forms, in the
axiomatic way, through its main properties listed below
\begin{eqnarray}
\nonumber
&&i_{\hat v}(\hat f)=0 \ , \\
\nonumber
&&i_{\hat \partial_i}(d\hat x^j)=\delta_i^{\,\,j} \ ,\\
&&i_{(\lambda\,\hat f\,\hat
\partial_i+\mu\,\hat g\,\hat
\partial_j)}(\hat\omega)=\lambda\,\hat f\,i_{\hat \partial_i}(\hat\omega)
+\mu\,\hat g\,i_{\hat \partial_j}(\hat\omega) \ ,\label{qc}\\
\nonumber &&i_{\hat v}\,(\lambda\,d\hat x^i\,\hat
f+\mu\,d\hat x^j\,\hat g)=
\lambda\,i_{\hat v}(d\hat x^i)\,\hat f+\mu\,i_{\hat v}(d\hat x^j)\,\hat g \ ,\\
\nonumber &&i_{\hat\partial_k}\,(d\hat x^1\wedge d\hat
x^2)=\delta_k^{\,\,1}\,d\hat
x^2-q^{-1}\,\delta_k^{\,\,2}\,d\hat x^1 \ ,
\end{eqnarray}
which are formally identical to Eq. (\ref{i}) with the
difference that now, the ordering is important. A rigorous
derivation of these properties can be found in
Ref. \cite{Aschieri}.\\
A realization of the above $q$-algebra and its $q$-calculus
can be accomplished by the replacements \cite{Ubriaco}
\begin{eqnarray}
&&\hat x\rightarrow x \ ,\label{x}\\
&&\hat p\rightarrow p\,D_x \ ,\label{p}\\
&&\hat\partial_x\rightarrow{\cal D}_x \ ,\label{dx}\\
&&\hat\partial_p\rightarrow{\cal D}_p\,D_x \ ,\label{dp}
\end{eqnarray}
where
\begin{equation}
D_x=q^{x\,\partial_x}\Rightarrow D_x f(x,\,p)=f(q\,x,\,p) \
,
\end{equation}
is the dilatation operator along the $x$ direction
(reducing to the identity operator for $q\rightarrow 1$),
whereas
\begin{eqnarray}
&&{\cal D}_x=\frac{q^{2\,x\,\partial_x}-1}{(q^2-1)\,x} \
,\\&&{\cal D}_p=\frac{q^{2\,p\,\partial_p}-1}{(q^2-1)\,p}\
,
\end{eqnarray}
are the Jackson derivatives (JD) with respect to $x$
and $p$ \cite{jack}. Their action on an arbitrary function $f(x,\,p)$ is
\begin{eqnarray}
&&{\cal
D}_x\,f(x,\,p)=\frac{f(q^2\,x,\,p)-f(x,\,p)}{(q^2-1)\,x} \
,\\
&&{\cal D}_p\,f(x,\,p)=\frac{f(x,\,q^2\,p)-f(x,\,p)}{(q^2-1)\,p} \,
\end{eqnarray}
which reduce to the ordinary derivatives when $q$ goes to
unity. Therefore, as a consequence of the non-commutative
structure of the $q$-plane, in this realization the $\hat
x$ coordinate becomes a $C$-number and its derivative is
the JD whereas the $\hat p$ coordinate and its derivative
are realized in terms of the dilatation operator and JD.

\section{Poisson bracket in the $q$-commutative phase space}

On the basis of the $q$-commutative differential calculus,
in this section we derive the expression for the $q$-deformed PB
by following, in analogy, the same formal steps used in the
classical derivation reviewed in Sec. 2. To begin with, let us introduce the
$q$-deformed symplectic form
\begin{equation}
\hat\omega=q^{-1/2}\,d\hat x\wedge d\hat p \ ,\label{qcsym}
\end{equation}
and we define the $q$-Hamiltonian field $\hat X_f$, associated
with the function $\hat f\in\widehat{\mathcal A}$, through the
relation
\begin{equation}
i_{\hat X_f}\,\hat\omega=d\hat f \ .
\end{equation}
According to  $q$-calculus, the expression for $\hat X_f$ is given
by
\begin{equation}
\hat X_f=q^{1/2}\,(\hat\partial_p\hat f)^{\rm
L}\,\hat\partial_x-q^{-1/2}\,(\hat\partial_x\hat f)^{\rm
L}\,\hat\partial_p \ ,\label{qham}
\end{equation}
which reduces to the standard Hamiltonian field (\ref{ham}) in the
$q\to1$ limit.

We introduce the $q$-Poisson bracket between the
$q$-defor\-med functions $\hat f$ and $\hat g$ by means of the
relation,
\begin{equation}
\Big\{\hat f,\,\hat g\Big\}_q\equiv i_{\hat X_{\hat
g}}(d\hat f)=i_{\hat X_{\hat g}}\,i_{\hat X_{\hat
f}}(\hat\omega) \ .\label{qpb}
\end{equation}
Accounting for Eq. (\ref{qd2}) and the properties (\ref{qc}) we
obtain the general expression for the $q$-PB
\begin{equation}
\Big\{\hat f,\,\hat g\Big\}_q\equiv
q^{1/2}\,(\hat\partial_p\hat g)^{\rm
L}\,(\hat\partial_x\hat f)^{\rm
R}-q^{-1/2}\,(\hat\partial_x \hat g)^{\rm
L}\,(\hat\partial_p\hat f)^{\rm R} \ .\label{qpb1}
\end{equation}
Properties of the $q$-PB (\ref{qpb}) are consequences of the
properties of the $q$-deformed contraction operator (\ref{qc}). In
particular they are bi-linear,
\begin{eqnarray}
&&\Big\{\lambda\,\hat f+\mu\,\hat g,\,\hat
h\Big\}_q=\lambda\,\Big\{\hat f,\,\hat
h\Big\}_q+\mu\,\Big\{\hat g,\,\hat h\Big\}_q \
,\\&&\Big\{\hat f,\,\lambda\,\hat g+\mu\,\hat
h\Big\}_q=\lambda\,\Big\{\hat f,\,\hat
g\Big\}_q+\mu\,\Big\{\hat f,\,\hat h\Big\}_q \ ,
\end{eqnarray}
but in general they are not skew-symmetric. This can be seen, for
instance, if we consider the $q$-deformed generator functions
which can be constructed by setting $\hat f\equiv \hat x$ and
$\hat g\equiv \hat p$. The canonical $q$-Hamiltonian fields are
respectively
\begin{eqnarray}
&&\hat X_x=-q^{-1/2}\,\hat\partial_p \ ,\\ &&\hat
X_p=q^{1/2}\,\hat\partial_x \ ,\label{qX}
\end{eqnarray}
and, after observing that, according to Eq. (\ref{der}),
$(\hat\partial_i\hat x^j)^{\rm L}\equiv (\hat\partial_i\hat
x^j)^{\rm R}$, from Eq. (\ref{qpb1}) we immediately derive
the $q$-de\-for\-med structure functions for the $q$-Poisson
algebra
\begin{equation}
\Big\{\hat x^i,\,\hat
x^j\Big\}_q=q^{1/2}\,\hat\partial_x\hat
x^i\,\hat\partial_p\,\hat x^j-q^{-1/2}\,\hat\partial_p\hat
x^i\,\hat\partial_x\,\hat x^j \ .\label{qpb3}
\end{equation}
Their explicit values are obtained as follows
\begin{eqnarray}
\nonumber
&&\Big\{\hat x,\,\hat x\Big\}_q=\Big\{\hat p,\,\hat p\Big\}_q=0 \ ,\\
&&\Big\{\hat x,\,\hat p\Big\}_q=q^{1/2} \ ,\label{qcan}\\
\nonumber&&\Big\{\hat p,\,\hat x\Big\}_q=-q^{-1/2} \ ,
\end{eqnarray}
so that we easily obtain
\begin{equation}
\Big\{\hat x,\,\hat p\Big\}_q=-q\,\Big\{\hat p,\,\hat
x\Big\}_q \ .\label{nonq}
\end{equation}
Finally, the Jacobi identity is trivially satisfied for the
$q$-structure functions
\begin{eqnarray}
\nonumber &&\Big\{\hat x^i,\,\Big\{\hat x^j,\,\hat
x^k\Big\}_q\Big\}_q+\Big\{\hat x^j,\,\Big\{\hat x^k,\,\hat
x^i\Big\}_q\Big\}_q\\+&&\Big\{\hat x^k,\,\Big\{\hat
x^i,\,\hat x^i\Big\}_q\Big\}_q=0 \ .
\end{eqnarray}

In Ref. \cite{Noi} a
$q$-deformed Poisson bracket has been obtained by requiring that,
in analogy with the classical case, where standard PB is invariant
with respect to a symplectic transformation, the $q$-PB should be
invariant with respect to a $q$-symplectic transformation. As a
result we obtained a $q$-PB formally equivalent to Eq.
(\ref{qpb1}) but with the replacement of $q\to1/q^2$. It is also
easy to see that the $q$-PB given in Eq. (\ref{qpb1}) is preserved
under the action of a $q$-symplectic transformation.
In order to show such a property, we recall
that, in the fundamental representation of Sp$_q$(1), any element
$\hat T$ is given by a $2\times2$ matrix, with entries $\hat
T_i^{\,\,j}$, satisfying the following equation \cite{Reshetikhin}
\begin{equation}
\hat T^{\,\,i}_r\,C^{rs}_q\,\hat T^{\,\,j}_s=C^{ij}_q \
,\label{ctt}
\end{equation}
where $C_q^{ij}=\epsilon^{ij}\,q^{-\epsilon^{ij}}$. Observing that
Eq. (\ref{qpb1}) can be expressed as
\begin{equation}
\Big\{\hat f,\,\hat g\Big\}_q=\left(\hat\partial_i\hat
g\right)^{\rm L}\,{\mathcal
J}_q^{ij}\,\left(\hat\partial_j\hat f\right)^{\rm R} \
,\label{qpoisson}
\end{equation}
where ${\mathcal J}_q=-C_{q^\prime}$, with $q^\prime=\sqrt{q}$,
according to the property (\ref{ctt}), it is easy to see that
under the action of quantum group Sp$_{q^\prime}$(1), Eq.
(\ref{qpoisson}) transforms as
\begin{equation}
\Big\{\hat f,\,\hat g\Big\}_q\to \Big\{\hat f^\prime,\,\hat
g^\prime\Big\}_q=\left(\hat\partial_i^\prime\hat
g^\prime\right)^{\rm L}\,{\mathcal
J}_q^{ij}\,\left(\hat\partial_j^\prime\hat
f^\prime\right)^{\rm R} \ ,
\end{equation}
where $\hat f^\prime\equiv \hat f(\hat x^\prime,\,\hat
p^\prime)$, $\hat g^\prime\equiv \hat g(\hat
x^\prime,\,\hat p^\prime)$, with
\begin{equation}
\hat x^i\rightarrow \hat x^{\prime i}=\hat x^j\,\hat
T_j^{\,\,i} \ ,\label{trasf}
\end{equation}
and we assumed the commutations between the group elements
and the plane elements.\\
It may be observed that Eq. (\ref{ctt}) can be rewritten as
\begin{equation}
\hat T^{\,\,i}_r\,{\mathcal J}^{rs}_q\,\hat
T^{\,\,j}_s={\mathcal J}^{ij}_q \ ,
\end{equation}
which mimics, in this way, the classical expression (\ref{cttc}),
where the matrix $\mathcal J_q$ plays the role of the symplectic
unit $J$ introduced in Eq. (\ref{j}) and recovered in the $q\to1$
limit. The $q$-deformed generator functions are then related to the
entries of the matrix $\mathcal J_q$ as
\begin{equation}
\Big\{\hat x^i,\,\hat x^j\Big\}_q=\hat\partial_r\hat
x^i\,{\mathcal J}_q^{rs}\,\hat\partial_s\,\hat x^j \
,\label{qgen}
\end{equation}
which also shows their invariance under the action of the
$q$-symplectic group Sp$_{q^\prime}$(1).


\section{$q$-deformed Hamilton's equations}

As a preliminary application of the $q$-PB derived in the previous
section, it is natural to investigate the effect due to the
$q$-commutativity of the coordinates of the phase space on the
time evolution of a classical object existing in this space. We
postulate that the dynamics in the deformed phase space is
described, in analogy with classical mechanics, by means of the
following $q$-deformed evolution equations written in the form
\begin{eqnarray}
&&\dot{\hat x}(t)=\Big\{\hat x(t),\,\hat H(\hat x,\,\hat
p)\Big\}_q \
,\label{hx}\\
&&\dot{\hat p}(t)=\Big\{\hat p(t),\,\hat H(\hat x,\,\hat
p)\Big\}_q \ .\label{hp}
\end{eqnarray}
It is assumed that time enters in the $q$-generators as a normal
parameter. The time derivative, indicated by a dot, means
$\dot{\hat x}=d\hat x/dt$ where $dt$ is a $C$-number. In this way,
the $q$-commutative algebra of $\dot{\hat x}(t)$ and $\dot{\hat
p}(t)$ and its $q$-calculus are the same as that of $d\hat x(t)$
and $d\hat p(t)$. $\hat H(\hat x,\,\hat p)$ is the $q$-Hamiltonian
function which is
assumed to  not depend explicitly on time.

According to Eq. (\ref{qpb1}) the evolution Eqs.
(\ref{hx}) and (\ref{hp}) can be written down in the form
\begin{eqnarray}
&&\dot{\hat x}(t)=q^{1/2}\,\left(\hat\partial_p\hat
H\right)^{\rm L} \
,\label{hx1}\\
&&\dot{\hat p}(t)=-q^{1/2}\,\left(\hat\partial_x\hat
H\right)^{\rm L} \ ,\label{hp1}
\end{eqnarray}
which are the $q$-deformed Hamilton's canonical
equations.

It can be shown that, as a consequence of Eqs.
(\ref{hx}) and (\ref{hp}), the time evolution of any dynamical
function \\
$\hat f(\hat x(t),\,\hat p(t);t)$, depending on the
generators $\hat x(t)$ and $\hat p(t)$, can be described by
means of the following evolution equation
\begin{equation}
\frac{d}{dt}\hat f(\hat x,\,\hat p;t)=\Big\{\hat f(\hat
x,\,\hat p;t),\,\hat H(\hat x,\,\hat
p)\Big\}_q+\frac{\partial}{\partial t}\hat f(\hat x,\,\hat
p;t) \ ,\label{evolution}
\end{equation}
where, the last term in the right hand side takes into account the
explicit dependence of $\hat f$ from $t$.\\ In fact, we recall
readily that the most general form of a function $\hat f\in\hat
{\mathcal A}$ can be written as a polynomial in the $q$-variables
$\hat x(t)$ and $\hat p(t)$
\begin{equation}
f(\hat x(t),\,\hat p(t);\,t)=\sum_{n,m}c_{nm}(t)\,[\hat
x(t)]^n\,[\hat p(t)]^m \ ,\label{f}
\end{equation}
where $c_{nm}(t)$ are $C$-numbers which may be time dependent
 and we have assumed the $\hat x$-$\hat p$ ordering
prescription which can be always accomplished by means of Eq.
(\ref{qplane}). Let us consider the generic term in Eq. (\ref{f}).
Its time derivative becomes
\begin{eqnarray}
\nonumber \frac{d}{dt}(c_{nm}\,\hat x^n\,\hat
p^m)&=&c_{nm}\,[n]_q\,\frac{d\hat x}{dt}\,\hat
x^{n-1}\,\hat p^m\\
\nonumber&+&c_{nm}\,[m]_q\,q^n\,\frac{d\hat p}{dt}\,\hat
x^n\,\hat p^{m-1}+\frac{\partial\,c_{nm}}{\partial t}\,\hat
x^n\,\hat p^m \ ,\\\label{mon}
\end{eqnarray}
where Eq. (\ref{qd}) has been employed and where we have
introduced the $q$-basic number
\begin{equation}
[n]_q=\frac{q^{2\,n}-1}{q^2-1} \ .
\end{equation}
By employing the  equations  of motion (\ref{hx}) and (\ref{hp})
in the form $d\hat x/dt=i_{\hat X_H}(d\hat x)$ and $d\hat
p/dt=i_{\hat X_H}(d\hat p)$, and accounting for the properties of
the operators $d$ and $i_{\hat v}$, we obtain
\begin{eqnarray}
\nonumber \frac{d}{dt}&&\hspace{-4mm}(c_{nm}\hat x^n\hat
p^m)=c_{nm}[n]i_{X_H}(d\hat x)\hat x^{n-1}\hat p^m\\
\nonumber& &+c_{nm}[m]q^ni_{X_H}(d\hat p)\hat x^n\hat
p^{m-1}+\frac{\partial c_{nm}}{\partial t}\hat x^n\hat
p^m\\
\nonumber &=&i_{X_H}\left(c_{nm}\left([n]d\hat x\hat
x^{n-1}\hat p^m\right.\right.\\ \nonumber&
&\left.\left.+[m]q^nd\hat p\hat x^n\hat
p^{m-1}\right)\right)+ \frac{\partial}{\partial
t}\left(c_{nm}\hat
x^n\hat p^m\right)\\
\nonumber &=&i_{X_H}\left(d(c_{nm}\hat x^n\hat
p^m)\right)+\frac{\partial}{\partial t}\left(c_{nm}\hat
x^n\hat p^m\right)\\
&=&\Big\{c_{nm}\hat x^n\hat p^m,\hat H(\hat p,\hat
x)\Big\}+\frac{\partial}{\partial t}\left(c_{nm}\hat
x^n\hat p^m\right) \ ,\label{proof}
\end{eqnarray}
which, by linearity, implies Eq. (\ref{evolution}).\\
Finally, we recall that, in  standard mechanics, the
canonical transformations are a kind of coordinate
transformations mixing $x$ and $p$ in a way that preserves
the Hamilton structure of the dynamical system. An
important property of the canonical transformations is that
they preserve the Poisson bracket. Thus, this family of
transformations, also called symplectic transformations
\cite{Abrahams}, are realized by the symplectic group
transformations. In the $q$-deformed case, the situation is
very similar because, as shown in the previous section, the
$q$-PB are invariant under the action the $q$-deformed
symplectic group. In this sense, the quantum group
Sp$_{q^\prime}$(1) is realized as a version of $q$-deformed
canonical transformation. It is outside
 the scope of this work to discuss  this important
subject of the theory and  its implications which will be
discussed in a future investigation.\\
In the following we are going to consider some non relativistic systems
described by the $q$-Hamiltonian
\begin{equation}
\hat H(\hat x,\,\hat p)=\frac{\hat p^2}{2\,m}+\hat V(\hat
x) \ ,\label{hamiltonian}
\end{equation}
where $m$ is an ordinary $C$-number and $\hat V(\hat x)$ is
the external potential which we assume to be an arbitrary
polynomial in the generator $\hat x$, with $C$-numbers
coefficients. Within this formalism let us discuss some
examples.

\subsection{The $q$-free particle}
As a first simple example, we choose $\hat V(\hat x)=0$, a free
particle. The $q$-Hamiltonian field is readily computed from Eq.
(\ref{qham}) and is given by
\begin{equation}
\hat X_{H_0}={\hat p\over m_q}\,\hat\partial_x \ ,
\end{equation}
where $\hat H_0=\hat p^2/2\,m$ and $m_q=2\,m\,q^{3/2}/[2]_q$.\\
The  equations of motion are obtained by employing Eq. (\ref{qpb})
and read
\begin{equation}
\dot{\hat x}={\hat p\over m_q} \ ,\hspace{20mm}\dot{\hat
p}=0 \ .
\end{equation}
These equations show that the effect of the deformation is to
rescale the mass $m$ of the particle as the effective mass $m_q$.
We remark that the $q$-Hamiltonian of the free particles $\hat
H_0$ is a constant of motion of the system. In fact, by using Eq.
(\ref{evolution}), it can be verified that $\{\hat H_0,\,\hat
H_0\}_q=0$ so that the Hamiltonian $\hat H_0$ could represent the
energy
of the free particle which is conserved in time.\\
This fact, in the undeformed canonical theory, is  merely a
consequence of the skew-symmetry of the PB:
$\{f,\,g\}=-\{g,\,f\}$ which implies $\{f,\,f\}=0$. In the
$q$-deformed case, because the $q$-PB is no longer
skew-symmetric, the $q$-PB evaluated between the same
function $\hat f$ in general does not vanish. An immediate
consequence of this is that, contrary to standard classical
mechanics, in the $q$-deformed theory, the Hamiltonian
function in general is not conserved in time. Such a
situation has been encountered also in other proposed
$q$-deformed classical systems \cite{Lukin}. Let us
investigate the consequence in the next example.

\subsection{The $q$-harmonic oscillator}

We consider the harmonic oscillator with a potential
\begin{equation}
\hat V_{\rm h}(\hat x)={1\over2}\,m\,\omega^2\,\hat x^2 \
,\label{ho}
\end{equation}
where the angular frequency $\omega$ is a $C$-number.\\
The $q$-Hamiltonian field is evaluated as
\begin{equation}
\hat X_{H_{\rm h}}={\hat p\over
m_q}\,\hat\partial_x-m_q\,{\omega_q}^2\,\hat
x\,\hat\partial_p \ ,
\end{equation}
where $\hat H_{\rm h}=\hat H_0+\hat V_{\rm h}$ and
$\omega_q=\omega\,[2]_q/2\,q^2$ while the  equations of
motion become
\begin{equation}
\dot{\hat x}={\hat p\over m_q} \ ,\hspace{20mm}\dot{\hat
p}=-m_q\,\omega_q^2\,\hat x \ .
\end{equation}
The effect of the deformation is thus taken into account only
through a rescaling of both the mass $m\to m_q$ and the angular
frequency $\omega\to \omega_q$ of the harmonic oscillator. The
time evolution of Hamiltonian $\hat H_{\rm h}$ can be determined
by employing Eq. (\ref{evolution}) and we obtain
\begin{equation}
\dot{\hat H}_{\rm h}=\Big\{\hat H_{\rm h},\,\hat
H_h\Big\}_q=q^{1/2}\,(q^2-1)\,\omega_q^2\,\hat p\,\hat x \
,
\end{equation}
which vanishes only in the $q\to1$ limit. As a consequence, the
Hamiltonian $\hat H_{\rm h}$ cannot be associated
with the total energy of the conservative system.\\
Nevertheless, the following function
\begin{equation}
{\mathcal E}_q=\frac{\hat
p^2}{2\,m}+{1\over2\,q^2}\,m\,\omega^2\,\hat x^2 \
,\label{energy}
\end{equation}
is time conserved,  $\dot{\mathcal E}_q=0$, and reduces, in the
$q\to1$ limit, to the Hamiltonian of the harmonic oscillator.
Hence, the non-conservation
of $\hat H$ is not necessarily fatal for the theory
due to the existence of other constants of  motions that could be
identified with the energy of the system instead of $\hat H$. In
other words, in the $q$-deformed theory, $\hat H$ is the generator
of the  equation of motion which now does not necessarily coincide with the
total energy of the system, but in correspondence, a new constant of
motion may be generated.

\subsection{The general case}

The previous result can be extended to the general case by
considering a $q$-deformed mechanical system governed by a
potential $\hat V(\hat x)$ defined by a polynomial series of $\hat
x$
\begin{equation}
\hat V(\hat x)=\sum_{n=1}c_n\,\hat x^n \ ,\label{pot}
\end{equation}
with $C$-number coefficients. The $q$-Hamiltonian field is
given by
\begin{equation}
\hat X_{H}={\hat p\over
m_q}\,\hat\partial_x-q^{-1/2}\,(\hat\partial_x\hat V)^{\rm
L}\,\hat\partial_p \ ,
\end{equation}
where $(\hat\partial_x\hat V)^{\rm L}\,d\hat x =d\hat
x\,\left(\hat\partial_x\hat V\right)^{\rm R}$, whereas the
equation of motion can be expressed in the form
\begin{equation}
m_q\,\ddot{\hat x}=\hat F_q \ ,
\end{equation}
with $\hat F_q=-q^{-1/2}\,(\hat\partial_x\hat V)^{\rm L}$
the $q$-deformed external force. Again, the Hamiltonian
is not conserved in time, whereas the function
\begin{equation}
{\mathcal E}_q=\frac{\hat p^2}{2\,m}+\sum_{n=1}d_n\,\hat
x^n \ ,\label{energy1}
\end{equation}
with $d_n=c_n\,q^{4-3\,n}$, reduces to the Hamiltonian
function of the system in the $q\to1$ limit and fulfills
the relation $\dot{\mathcal E}_q=0$.


\section{Canonical quantization}

Let us now compare, by means of canonical quantization, the
well-known $q$-deformed Heisenberg algebra with the algebra
(\ref{qcan}) generated by the $q$-deformed PB. To start with we
recall that the $q$-deformed Heisenberg algebra reads
\cite{Hinterding,Cerchiai,meljanac,Zhang}
\begin{eqnarray}
&&\Big[\tilde x,\,\tilde p\Big]_q=i\,\tilde\Lambda_q \ ,\label{qc1}\\
&&\tilde\Lambda_q\,\tilde x=q^{-1}\,\tilde x\,\tilde \Lambda_q \
,\hspace{10mm}\tilde\Lambda_q\,\tilde p=q\,\tilde p\,\tilde
\Lambda_q \ ,\label{qc2}
\end{eqnarray}
where
\begin{equation}
\Big[\tilde x,\,\tilde p\Big]_q=q^{1/2}\,\tilde x\,\tilde
p-q^{-1/2}\,\tilde p\,\tilde x \ ,\label{qdc}
\end{equation}
is the $q$-deformed commutator and we have denoted by a tilde the
$q$-deformed quantum generators $\tilde x$ and $\tilde p$. The extra
generator $\tilde\Lambda_q$ in Eqs. (\ref{qc1}) and (\ref{qc2})
plays the role of a dilatator and, in the $q\to1$ limit, reduces
to the identity operator \cite{Cerchiai,Wess1}.\\
As it is known, in the undeformed case, the simplest canonical
quantization prescription is given by identify the position variable
$x$ with the corresponding multiplicative operator $\overline x$ and
the canonically conjugate variables $p$ is assumed to be
proportional to the space derivative according to the rule
\begin{equation}
x \rightarrow \overline x \ ,\hspace{20mm} p \rightarrow \overline
p\equiv -i\,\partial_x \ .
\end{equation}
At the same time, the undeformed PB is replaced with the
undeformed commutator
\begin{equation}
\Big\{x,\,p\Big\}=1 \Leftrightarrow \Big[\overline x,\overline
p\Big]=i \ ,
\end{equation}
where $[\overline x,\,\overline p]=\overline x\,\overline
p-\overline p\,\overline x$. In analogy with this scenario, it
appears natural to impose, in the $q$-deformed framework, the
following quantization rule on the $q$-variables
\begin{eqnarray}
&&\hat x \rightarrow \tilde X \ ,\label{p1}\\ && \hat p \rightarrow
\tilde P\equiv-i\,\hat\partial_{\tilde X} \ ,\label{p2}
\end{eqnarray}
and to replace the $q$-PB with the $q$-deformed commutator
(\ref{qdc}).\\ It should be observed that, contrary to the
$q$-deformed classical theory developed starting from the
2-dimensional $q$-deformed calculus, the $q$-deformed quantum
mechanics is spanned in the 1-dimensional configuration space.
This requires us to consider the 1-dimensional $q$-deformed
calculus generated by the elements $\hat x,\,\hat\partial_x$ and
$d\hat x$ which differs from the corresponding one in 2-dimensions
and introduced in Section 3. In particular, the Leibnitz rule now
reads \cite{Wess}
\begin{equation}
\hat\partial_x\hat x=1+q\,\hat x\,\hat\partial_x \ ,\label{qpl}
\end{equation}
which differs from the last of Eq. (\ref{qco}).\\
Let us observe that, due to Eq. (\ref{qpl}), assuming $\tilde X$
is a Hermitian quantity, the prescription (\ref{p2}) deals with a
non Hermitian definition of momentum. In accordance with current
literature, we can define a physical momentum
\begin{equation}
\tilde p={1\over2}\,\left(\tilde P+\tilde P^\dag\right) \
,\label{pp}
\end{equation}
where the Hermitian conjugation of $\hat\partial_X$ is defined by
\begin{equation}
\hat \partial_{\tilde X}^\dag=-q^{-1/2}\,\tilde
\Lambda_q\,\hat\partial_{\tilde X} \ ,
\end{equation}
and the unitary operator $\tilde \Lambda_q^\dag=\tilde
\Lambda_q^{-1}$ takes the expression \cite{Cerchiai,Wess1}
\begin{equation}
\tilde \Lambda_q=q^{-1/2}\,\Big[1+(q-1)\hat X\,\hat\partial_{\tilde
X}\Big]^{-1} \ .\label{lambda}
\end{equation}
After the redefinition
\begin{equation}
\tilde x={2\,q\over1+q}\,\tilde X \ ,\label{xp}
\end{equation}
it is easy to verify that the definitions (\ref{pp}), (\ref{lambda}) and
(\ref{xp}) fulfils Eqs. (\ref{qc1}) and Eq. (\ref{qc2}). Thus, by
posing formally
\begin{equation}
\hat x \rightarrow \tilde x \ ,\hspace{10mm} \hat p \rightarrow
\tilde p \ ,\hspace{10mm}\sqrt{q}\rightarrow i\,\tilde\Lambda_q \
,\label{corresp}
\end{equation}
we can state the following correspondence
\begin{equation}
\Big\{\hat x,\,\hat p\Big\}_q=\sqrt{q} \Leftrightarrow \Big[\tilde
x,\,\tilde p\Big]_q=i\,\tilde\Lambda_q \ ,\label{qqop}
\end{equation}
among the $q$-PB between the classical space phase variables and the
$q$-commutator between the corresponding quantum operators.\\
Finally, by taking the Hermitian conjugate of Eq. (\ref{qc1}),
accounting for the unitariety of $\tilde\Lambda_q$, we obtain
\begin{equation}
\Big[\tilde p,\,\tilde x\Big]_q=-i\,\tilde\Lambda_q^{-1} \
.\label{her}
\end{equation}
We can verify that Eqs. (\ref{qqop}) and (\ref{her}) are completely
consistent with Eqs. (\ref{qcan}), under the correspondence
(\ref{corresp}). Equivalently, we can verify that the relation
\begin{equation}
\Big\{\hat x,\,\hat p\Big\}_q=-\Big\{\hat x,\,\hat p\Big\}_q^{-1} \
,
\end{equation}
which follows directly from Eqs. (\ref{qcan}), is replaced by
\begin{equation}
\Big[\tilde x,\,\tilde p\Big]_q=-\Big[\tilde x,\,\tilde
p\Big]_q^{-1} \ ,
\end{equation}
as can be verified by a direct calculation.\\
In  light of the above prescriptions, starting from the classical
evolution equation (\ref{evolution}), it appears natural to
postulate the $q$-deformed Heisenberg equation for an operator
$\tilde O_q$ as follows
\begin{equation}
\frac{\partial \tilde O_q}{\partial t}=i\,[\tilde O_q,\,\tilde H]_q
+\frac{\partial \tilde O_q}{\partial t} \ .
\end{equation}


\section{Conclusion}

In this paper, we have developed a $q$-deformed version of the
Poisson bracket by generalizing, in the quantum groups framework,
the method based on the exterior calculus for the definition of
the classical PB. We have derived, in a systematic way, an
expression of the $q$-PB which substantially equivalent to the one
recently obtained by us in Ref. \cite{Noi} by following a different
and more consistent method.
The two approaches differ in the following sense. In the
previous one \cite{Noi}, the expression for the $q$-PB was
conjectured by requiring the invariance of the $q$-PB under the
action of the symplectic group Sp$_q$(1), leading to the
$q$-deformation of the phase space,  whereas in the present
investigation, only the non-commutative $q$-deformation of the phase space has
been imposed and the $q$-PB has been obtained as a consequence of
the $q$-calculus. It has been shown that the new version of the
deformed bracket is still invariant under the action of the
$q$-symplectic group of transformations Sl$_{q^\prime}$(2) with
$q^\prime=\sqrt{q}$.

We have discussed some properties of the $q$-deformed PB
and we have presented some elementary examples to
illustrate how a possible $q$-deformed classical mechanics
can be introduced. It has been shown that, in contrast to
the undeformed case, the $q$-Hamiltonian, which is the
generator of the evolution equation, is in general not
conserved in time and cannot be identified with the total
energy for conservative systems. However, a suitable
function, reducing to the Hamiltonian function in the
$q\to1$ limit and remaining constant during the evolution
of the system, has been obtained for the examples studied.
These properties, related to derivation of the $q$-PB given
in Eq. (\ref{qpb1}), are the most relevant results of this
paper.

Finally, we have discussed a possible quantization me\-thod in the
$q$-deformed picture. Based on the standard method consisting of the
replacement of the (undeformed) PB for canonically conjugate
variables with the (undeformed) commutator for the corresponding
quantum operator, we have postulated a similar scheme by replacing
the $q$-PB for quantum conjugate variables with the $q$-deformed
commutator of the corresponding $q$-deformed quantum operators.

In conclusion we would like to mention the possible applications
of the $q$-deformed classical mechanics that we have developed to
the study of some relevant physical phenomenologies and in
particular to the framework of thermostatistics  \cite{pre} where
it could lead to a generalization of the theory in a manner
similar to what the classical Tsallis' thermostatistics does with
respect to the Boltzmann-Gibbs theory \cite{gellmann}.


\end{document}